\documentclass[12pt,a4paper]{article}
\usepackage{a4wide}
\usepackage{latexsym}
\usepackage{epsf}
\usepackage{amssymb}
\begin{document}
\def\be{\begin{equation}}
\def\ee{\end{equation}}
\def\bea{\begin{eqnarray}}
\def\eea{\end{eqnarray}}

\def\pd{\partial}
\def\a{\alpha}
\def\b{\beta}
\def\g{\gamma}
\def\d{\delta}
\def\m{\mu}
\def\n{\nu}
\def\t{\tau}
\def\l{\lambda}

\def\s{\sigma}
\def\e{\epsilon}
\def\scri{\mathcal{J}}
\def\cM{\mathcal{M}}
\def\tcM{\tilde{\mathcal{M}}}
\def\RR{\mathbb{R}}

\hyphenation{re-pa-ra-me-tri-za-tion}
\hyphenation{trans-for-ma-tions}


\begin{flushright}
IFT-UAM/CSIC-02-05\\
hep-th/0205075\\
\end{flushright}

\vspace{1cm}

\begin{center}

{\bf\Large Rudiments of Holography}

\vspace{.5cm}

{\bf Enrique \'Alvarez, Jorge Conde and Lorenzo Hern\'andez }

\vspace{.3cm}

\vskip 0.4cm

{\it  Instituto de F\'{\i}sica Te\'orica UAM/CSIC, C-XVI,
and  Departamento de F\'{\i}sica Te\'orica, C-XI,\\
  Universidad Aut\'onoma de Madrid 
  E-28049-Madrid, Spain }

\vskip 0.2cm

\vskip 1cm

{\bf Abstract}

An elementary introduction to Maldacena's AdS/CFT correspondence is given,
with some emphasis in the Fefferman-Graham construction. This is based on
lectures given by one of us (E.A.) at the Universidad Autonoma de Madrid.

\end{center}

\begin{quote}

\end{quote}


\newpage

\setcounter{page}{1}
\setcounter{footnote}{1}
\tableofcontents
\newpage
\section{Introduction}
The following is an introduction to the AdS/CFT correspondence proposed by
Maldacena in \cite{malda}.
\par
The style is informal, and only some computations towards the end are done in
detail. The topic is already large, and a comprehensive (up to the date it was
written) review is available (\cite{aharony}). References are only given to
material which has actually used in the original lectures, and are not complete in any
sense.
\par
Although strings almost do not appear at the level of approximation we shall
work, they lurk in the horizon.
Standard introductions to strings are \cite{green},\cite{polchinski} 
(cf. also \cite{alvarezmeessen} for a quick overview).
\par
We shall use the Landau-Lifshitz Timelike Conventions for General Relativity;
 that is, metric
signature $s=-2$, the Riemann tensor defined as $
R^{\a}\,_{\b\gamma\delta}=\pd_{\gamma}\Gamma^{\a}_{\beta\delta}-\ldots$, and
the Ricci tensor defined by $R_{\a\b}\equiv R^{\gamma}\,_{\a\gamma\b}$.
The flat line element with $p$ times and $q$ spaces will be denoted by
\be
d\vec{x}^2_{(p,q)}\equiv\sum_1^p (d x^i)^2 - \sum_1^q (d x^i)^2
\ee
Volume elements will be similarly shortened:
\be
dx_d\equiv \sqrt{g}dx^1\wedge\ldots\wedge dx^d
\ee
\newpage
\section{The Holographic Principe}
The line of reasoning that led G. 't Hooft to propose the {\em holographic
  principle} (\cite{hooft}\cite{bigatti}) stems from noticing that if we want
  a piece of matter of given mass $M$, and contained in a given volume
  $V\equiv L^3$
  to be observable from far away, we have to assume that the volume considered
  is bigger
  than the Schwarzschild scale:
\be
R_s\leq L
\ee
where the Schwarzschild radius is given by:
\be
R_s\equiv\frac{2 G M}{c^2}
\ee
because otherwise the system as a whole would be the interior of a black 
hole and, as such, could not be observed from the outside.
This means that if we assume the system to be at equilibrium, then, in first
  approximation at high enough temperature, the energy is given by
\be
E\sim V T^4
\ee
and the corresponding entropy is
\be
S\sim V T^3
\ee
and the bound just espoused is equivalent to
\be
S\leq \frac{V^{1/2}}{G^{3/4}}\sim (\frac{A}{G})^{3/4}
\ee
which is still quite small compared with the black hole entropy
\be
S=\frac{1}{4} A
\ee
It seems strange at first sight that the entropy is not proportional to the
  volume
\be
S\sim V
\ee
as would have been predicted by ordinary quantum field theory. One has to
  conclude that most of the would be quantum field states lie inside their 
 own Schwarzschild radius.
For consistency one is then led following this line of reasoning to the  postulate that the quantum theory of gravity
should be descibed by a sort of topological quantum theory, in the sense that
  all degrees of freedom could be projected onto the boundary.
\newpage
\section{The Maldacena conjecture}
\subsection{Physics on the world volume of a D-brane versus 
the brane as a source of spacetime curvature}
In the last few years following the seminal work of Polchinski 
(\cite{polchinski}) it has became increasingly clear that some topological
 defects, spanning a given number $p$ of space dimensions, generically called 
{\em branes},
play an essential r\^ole in the formulation of string theory. The simplest 
ones are those which can be defined as the locus of open string endpoints, 
which 
are called $D$(irichlet)-branes. 
\par
There is a fascinating duality between two different aspects of the
physics of D-branes (\cite{klebanov}): the physics on the brane, 
described by the Dirac-Born-Infeld (DBI) action, and the spacetime
 physics stemming from the brane as a source of energy-momentum.
\par
We indeed know (\cite{leigh})
 that the effective action of a D-brane,
(from which the Weyl anomaly coefficients are derived by a variational
principle) is the DBI action, which in the simplest
case reads:
\be
S_p \equiv - T_p \int d^{p+1}\xi e^{-\Phi} 
\sqrt{det[g_{ab}+b_{ab}+2\pi\a' F_{ab}]}
\ee
where $T_p$ is the brane tension, of mass dimension $p+1$,  and 
$g_{ab}\equiv g_{\m\n}\pd_a x^{\m}\pd_b x^{\n}$ is the metric
induced on the brane by the imbedding from the world volume of the brane, 
$\Sigma$, into the external spacetime, $M$,
\be
\xi^a\in \Sigma_{p+1}\rightarrow x^{\m}\in M_{10}
\ee
(and similarly for $b_{ab}$).The gauge field strength is not 
pulled back from
spacetime; gauge degrees of freedom live on the brane only.
\par
On the other hand, it can be argued \cite{ortin}
that the gravitational field 
produced by a black D-brane of $p$ spacelike dimensions is given 
(in the string frame) by the universal formula
\footnote{When there are no extra translational isometries}:
\bea
&&ds^2= H^{-1/2}(r)[W(r) dt^2 - \d_{ij}dx^i dx^j]\nonumber\\
&& - H^{1/2}(r)[W^{-1}(r) dr^2 + r^2 d\Omega_{8-p}^2]
\eea
with $i,j,\ldots=1,\ldots p$
and a dilaton field given by
\be
e^{\phi - \phi_0} = H^{\frac{3-p}{4}}(r)
\ee
the functions $H$ and $W$ are given by:
\be
H(r)\equiv 1 + \frac{l^{7-p}}{r^{7-p}}
\ee
and
\be
W(r)\equiv 1 - \frac{r_0^{7-p}}{r^{7-p}}
\ee
and $d\Omega_{8-p}^2$ is the line element on the sphere $S^{8-p}$.
The source is at $r = r_0$, and the extremal limit is $r_0 = 0$, that is,
 $W = 1$.
\par
There is a RR background as well, given by the $p+1$-form:
\be
A_{p+1}\equiv \a e^{\phi_0}(H^{-1}-1)H^{-1/4}W^{1/2}dt\wedge dx^1\wedge
\ldots dx^p
\ee
and a relationship between the constants, namely:
\be
r_0^{7-p}=l^{7-p}(1-\a^2)
\ee
\par
Let us now concentrate, for the rest of this work, in the case $p=3$,
in which the world volume is four-dimensional and, besides, the dilaton
is constant.
\par
%
%

The self-dual RR five-form field strength is actually given by 
\be F_5\equiv 4 l^4 (\epsilon_5
+ \star \epsilon_5) 
\ee 
$\epsilon_n$ being the volume element on $S^n$.  
\par 
The
normalization is done as follows \cite{petersen}.  We first define the
charge of the brane, 
\be 
\mu_3\equiv
\frac{1}{\sqrt{2\kappa_{10}^2}}\int_{S_5}\star F_5 
\ee 
which yields
$\mu_3 = \frac{4 l^4 \Omega_5}{\sqrt{2\kappa_{10}^2}}$.  
The BPS condition
(in the case in which we have $N$ coincident branes; that is, a BPS
system of charge $N$), is the equivalent to the following relationship
with the string tension: 
\be \mu_3 = N\sqrt{2\kappa_{10}^2}\tau_3 
\ee
which leads to (after plugging the values $\tau_3=\frac{1}{g_s
(2\pi)^3\alpha'^2}$ and $\kappa_{10}^2=2^6\pi^7\alpha'^4 g_s ^2$ taken from
\cite{polchinski}) 
\be l^4= 4 \pi g_s N l_s^4 
\ee 
where $\alpha'\equiv l_s^2$.

\subsection{Absorption Cross Sections}
The fate of scalar particles when approaching a D-brane can be computed 
from the two different viewpoints alluded above.
\par
Indeed I. Klebanov (\cite{klebanov}) realized 
that absorption cross sections
could be calculated both from the D-brane (DBI) point of view, or from the
gravitational field of the D-brane itself, with identical results.
\par
In the simplest case of a dilaton incident at right angles with frequency 
$\omega$ the relevant DBI coupling (obtained from an low energy, weak field 
expansion of the full DBI action) is:
\be
S_{int} = \frac{\pi^{1/2}}{\kappa_{10}}\int d^4 x \frac{1}{4} \Phi\,tr F_{\m\n}^2
\ee
This implies that the cross section for decaying into a pair of gluons 
with vanishing spacial momentum, is given by:
\be
\sigma_{DBI} = \frac{\kappa_{10} ^2 \omega^3 N^2}{32\pi}
\ee
(where the factor $N^2$ comes from the degeneracy of the final state).
\par
On the other hand, the radial part of the equation of motion for a dilaton 
$\Phi(x)\equiv \phi(r) e^{i\omega t}$ is just
\be
[\frac{1}{r^5}\frac{d}{dr}r^5 \frac{d}{dr} + 
\omega^2 (1 + \frac{l^4}{r^4})]\phi(r) = 0
\ee
Using the covenient variable in the inner region,
\be
z\equiv \frac{\omega l^2}{r}
\ee
and substituting
\be
\phi(r) = z^{3/2} f(z)
\ee
yields
\be
(\frac{d^2}{dz^2} - \frac{15}{4z^2} + 1 + \frac{(\omega l)^4}{z^4}) f = 0
\ee
This implies an absorption cross section:
\be
\sigma_{sugra} = \frac{\pi^4}{8} \omega^3 l^8
\ee
Both cross sections can be shown to coincide by using the relationship
\footnote{ This is nothing more than the consistency condition
equating the  mass of the extremal D-brane solution with RR charge
with N times the mass of a single brane}
\be
l^4 = \frac{\kappa_{10}}{2\pi^{5/2}}N \sim g_s l_s^4 N
\ee
The range of validity of the supergravity calculation is 
\be
l>>l_s \Leftrightarrow Ng_s>>1
\ee
The condition that the incident energy is small is
\be
\omega l_s <<1
\ee
On the gauge theory side, this corresponds (because $g_{YM}^2\sim g_s$) to 
large 't Hooft coupling, $g^2_{YM} N\rightarrow 0$. If we want to supress
string loop corrections, we need in addition $g_s\rightarrow 0$, implying that
$N>>1$.
\subsection{Maldacena's near horizon limit}

Motivated by the preceding results, Maldacena (\cite{maldacena})
realized that when we are simultaneously interested in the {\em near horizon}
solution (which means $r\rightarrow 0$) and low energies
(that is, $\a'\rightarrow 0$), there is a natural variable that can 
be introduced (a natural {\em blow up}, from the mathematical point of view),
namely:
\be
u\equiv\frac{r}{\a'}
\ee
This variable has dimensions of energy, and, in spite of the fact that
the starting point is the near horizon limit, the variable clearly 
is a continuus one, and it can reach arbitrary real values.
\par
It has indeed been suggested that this variable has 
to do with the renormalization group scale of the gauge 
theory living on the stack of branes, namely $\mathcal{N}=4$ 
SUSY Yang Mills.
\par
Performing the limit in the supergravity solution, and using
$l^4 = 4\pi g_s N \a'^2$ yields
\be
ds^2 = \a'[ \frac{u^2}{\sqrt{4\pi g_s N}} dx_{\parallel}^2 + 
\sqrt{4\pi g_s N}\frac{du^2}{u^2} + \sqrt{4\pi g_s N}d\Omega_5^2]
\ee
which happens to be the metric of Anti de Sitter spacetime of radius 
$l^2=l_s^2 \sqrt{4\pi g_s N} $
cross a five sphere of the same radius, $AdS_5\times S_5$.(\cite{gibbonstownsend}).
\par
And the physics on the brane itself is described by 
a $d=4$ conformal field theory (CFT),
 namely $\mathcal{N}=4$ SUSY Yang-Mills, with gauge group $SU(N)$ and coupling
 constant $g=g_s^{1/2}$.
The results of the last section lead us to espect that there
is a close relationship between these two descriptions of the D-brane stack.
\par
The lagrangian of 
$N=4$ supersymmetric Yang-Mills in 4 dimensions is
 given by
\bea
L &=& -\frac{1}{4}Tr\left( F_{\m\n}F^{\m\n}\right) 
+i\lambda_i\sigma^{\m}D_{\m} \bar{\lambda}^i 
+\frac{1}{2}D_{\m}\Phi_{ij}D^{\m}\Phi^{ij} \nonumber\\
&& + i\lambda_i 
[\lambda_j ,\Phi^{ij}] + i \bar{\lambda}^{i}[\bar{\lambda}^j ,\Phi_{ij}] +
 \frac{1}{4}[\Phi_{ij},\Phi_{kl}][\Phi^{ij},\Phi^{kl}] \; .
\eea
where the gauginos are represented by 
four Weyl spinors $\lambda_i$, transforming in the $\mathbf{4}$ of $SO(6)$,
and the six scalar fields $\Phi_{ij}$ obey 
$(\Phi_{ij})^{\dagger}\equiv \Phi^{ij} = 
\frac{1}{2}\epsilon^{ijkl}\Phi_{kl}$. Sometimes we shall represent the six
scalars as $\Phi^I$.
\par

The ten-dimensional Newton's constant is given by:
\be
\kappa_{10}^2\sim l_p^8 = g_s ^2 l_s^8 \sim \frac{l^8}{N^2}
\ee
The effective string tension is just
\be
T_{eff}\equiv\frac{l^2}{l_s^2}\sim \sqrt{g_s N}\sim \lambda^{1/2}
\ee
where we have introduced the 't Hooft coupling $\lambda\equiv g^2 N$.
The 't Hooft limit is precisely
\bea
&&g\rightarrow 0\nonumber\\
&&N\rightarrow\infty\nonumber\\
&&\lambda\equiv g^2 N\rightarrow constant
\eea
and corresponds to quasi-free strings. There is a slightly different, more holographic limit, to wit
\bea
&&g \rightarrow constant\\
&&N\rightarrow\infty\nonumber\\
&&l/l_s\rightarrow\infty
\eea
In this limit the effective string tension grows large, so that it is to be
expected that classical supergravity is a good approximation.
\par
Actually, from this starting point
a whole {\em mapping} (\cite{witten}) between strings on one side and CFT on the other has been slowly inferred. Indeed, from
this point on, one can forget about the way the conjecture was first posited,
and consider $AdS_5\times S_5$ as a new string background by itself. 
These considerations do allow the calculation
(in the aforemetioned large $N$, large 't Hooft coupling)
of gauge invariant correlators in the gauge theory side, using
{\em tree level} supergravity computations; that is, computing the action
of supergravity of certain {\em bulk} fields $\Phi_i$ with given values in the
(conformal) boundary, $\Phi_i|_{\pd}=\phi_i$. The mapping itself is the
association $\phi_i\rightarrow \mathcal{O}_i$, obtained through an expansion
of the DBI action.
\par
The whole setup is summarized in the generating functional (\cite{witten})
\be\label{generating.functional}
<e^{\int\sum \mathcal{O}_i \phi_i}> 
= e^{- S_{sugra}[\Phi_i]}|_{\Phi_i|_{\pd}=\phi_i}
\ee
It is plain that if the operators $\mathcal{O}_i$ have conformal weight
$\Delta_i$, then the associated fields $\phi_1$ (which in practice 
behave as currents
on the boundary) will have conformal dimension $4 -\Delta_i$.
\par
Maldacena actually proposed that for any value of the coupling there was
an exact quantum equivalence, between $IIB$ string theory in $AdS_5\times S_5$
and $\mathcal{N}=4$ SUSY Yang-Mills, for any value of $G=SU(N)$. 
From the string theory side, the value of $N$ appears as the Ramond-Ramond
flux on the sphere $S^5$.
\par
The global symmetries of both theories are the same, namely  $SO(4,2)\times
 SO(6)$,which includes the four dimensional  conformal
group which appears as an isometry group from the
string side, and as a $R$-symmetry $SU(4)\sim SO(6) $ in the CFT side. When 
fermions 
are considered,
both groups appear as the bosonic part of the supergroup $SU(2,2|4)$.
Besides a non perturbative S-duality group $SL(2,\mathbb{Z})$ is conjectured
 to exist on 
both sides.
\subsection{The Infrared/Ultraviolet Connection}
Anti de-Sitter space is non-compact; its volume $Vol(AdS_p)$ diverges. An
infrared (IR) regulator in the bulk (such as making believe that the boundary
is at $r=\epsilon$ instead of $r=1$ in the form (\ref{sw}) of the AdS metric
to be introduced momentarily) is
equivalent to an ordinary ultraviolet (UV) cutoff in the CFT living on the
boundary. Giving the fact that in the gauge theory  there are  in the large N
limit approximately $N^2$ degrees
of freedom per point, the number of degrees of freedom per unit of
three-dimensional volume
in the cutoff theory will be:
 \be
N_{d.o.f.}=\frac{N^2}{\epsilon^3}
\ee
Now the regularized area of the eight-dimensional spatial boundary at constant
time  is (taking into account an $l^5$
factor from the sphere $S^5$)
\be
A=\frac{l^8}{\epsilon^3}
\ee
in such a way that the number of degrees of freedom per unit boundary area is
\be
\frac{N_{d.o.f.}}{A}=\frac{N^2}{l^8}=\frac{1}{G}
\ee
in accordance with the holographic principle.
\newpage
\section{Structure of the Anti de Sitter Geometry. }

Given the basic importance of Anti de Sitter metric in the whole 
description of the spacetime region close to the brane,
 we have collected here some geometric facts, relevant for the 
discussion of boundary conditions in the main text, specially in connection
with the generating functional formerly introduced 
in the equation (\ref{generating.functional}).

Anti de Sitter space in p dimensions ($AdS_p$) is the symmetric space
\be
AdS_p\equiv SO(2,p-1)/SO(1,p-1)
\ee
Indeed, all real forms of $SO(p+1)/SO(p)$ are closely related through 
Weyl's unitary trick. This suggests the definition of an {\em euclidean} version
of AdS:
\be
EAdS_p\equiv SO(1,p)/SO(p)
\ee

\par
$AdS_p$ could also be explicitly defined
(\cite{gibbons}) as the induced metric on the hyperboloid 
\be
(X^0)^2 +
(X^p)^2 - \d_{ij} X^i X^j = l^2;
\ee
$ (i,j = 1\ldots p-1)$ embedded in an
ambient space $\mathbb{R}_{2,p-1}$ (that is, $\mathbb{R}_{p+1}$
endowed with a Minkowskian metric with two times, 
\be
ds^2 = (dX^0 )^2 +
(dX^p )^2 -\d_{ij} dX^i dX^j .
\ee
 Defined in that way, it clearly has
topology $S^1\times \mathbb{R}^{p-1}$ (as well as closed timelike
curves).  The universal covering space ($CAdS_p$) has topology
$\mathbb{R}^p$.
\par
$AdS$ is an Einstein space; its Ricci tensor is proportional to the
metric: 
\be 
R_{\m\n}=\frac{p-1}{l^2}g_{\m\n} 
\ee 
which corresponds to a
{\em positive} \footnote{The unconventional sign for the {\em AdS} 
cosmological constant is due to our choice of signature.} 
cosmological constant, 
\be 
\lambda = \frac{(p-1)(p-2)}{2 l^2} 
\ee
\par

This definition makes it manifest the underlying $O(2,p-1)$ symmetry.
The $p(p+1)/2$ Killling vectors are given by
\be
k_{ab}\equiv X^a\pd_b - X^b\pd_a
\ee
(for $0< a,b < p$).
As it is well known, there is a $2-1$ correspondence between
$O(2,p-1)$ and the conformal group of Minkowski space in $p-1$
dimensions, $C(1,p-2)$.

A first, provisional, definition of the {\em boundary} at infinity 
$\partial AdS$ can be defined as the
region where all $X^{\m}$ are rescaled by an infinite amount,
$X^{\m}\rightarrow \xi X^{\m}$, where $\xi\rightarrow \infty$ .
In that way, the boundary is characterized by the relationship 
$(X^0)^2 + (X^p)^2 - \d_{ij} X^i X^j = 0$, which is nothing but the well-known
$O(2,p-1)$ null-cone compactification of Minkowski space, $M^{\mathbb{C}}$.
\cite{penrose}.The way it works 
is that to any regular point of Minkowski space,$x^{\m}\in M$, there
corresponds another point in $M^{\mathbb{C}}$, namely 

\begin{equation}
\left\{  
\begin{array}{rcl}
X^0 &=& x^0\, ,\\
& & \\
X^i &=& x^i\, ,\\
& & \\
X^{p-1} &=& \frac{1+ x^2}{2}\, ,\\
& & \\
X^p &=& \frac{1 - x^2}{2}\, .\\
\end{array}
\right.
\end{equation}

The points in $M^{\mathbb{C}}$ which are not in $M$ correspond to $X^p
+ X^{p-1} = 0$. This means that this compactification amounts to add
an extra null cone at infinity.
\par
The $AdS$ metric can be easily put in the {\em globally static form}
by means of the ansatz (in which we introduce two closely related sets
of coordinates simultaneously)

\begin{equation}
\left\{
\begin{array}{rcl} 
X^0 &=& l \cos \t \cosh \chi\ = l \frac{cos \t}{cos\rho}, ,\\
& & \\
X^p &=& l \sin \t \cosh \chi\ = l \frac{sin\t}{cos\rho}, ,\\
& & \\
X^i &=& l~ n^i~\sinh \chi = l\, n^i\, \tan{\rho}\, ,\\
\end{array}
\right.
\end{equation}

\noindent where $\d_{ij}n^i n^j = 1, (i,j = 1,\ldots,p-1)$. The result , 
in terms of the first set of coordinates, is

\begin{equation}
ds^2 = l^2[(\cosh \chi)^2 d\t^2 - (d\chi)^2 - 
(\sinh \chi)^2 d\Omega_{p-2}^2 ]\, .
\end{equation}

\noindent $AdS$ corresponds to $0 \leq \t \leq 2\pi$, and $CAdS$ to 
$0 \leq \t \leq \infty$. The boundary lies at $\chi=\infty$.

The antipodal map $ J: X\rightarrow - X$, corresponds in this
coordinates simply to $ (\t,\chi,\vec{n}) \rightarrow (\t + \pi,\chi,
- \vec{n})$.
\par
 The second set of equalities gives the form conformal
to Einstein's static universe as used in \cite{avis}, 
\be\label{avis}
ds^2 =\frac{l^2}{cos^2 \rho}[d\tau^2 - d\rho^2 - \sin{\rho}^2\, d\Omega_{p-2}^2]
\ee
where $0\leq\rho<\pi/2$,$0\leq\theta<\pi$ and $0\leq\phi<2\pi$. The boundary is
now located at $\rho=\pi/2$. This clearly shows that the boundary is {\em
  timelike}, because the normal vector is spacelike. This fact is the root of
much of the peculiar behavior of this space.
\par
Hawking and Page (\cite{hawkingpage}) parametrize this as
\be
r\equiv l\, \tan{\rho}
\ee
and
\be
T\equiv l\tau \sim T + 2\pi l
\ee
in such a way that the metric reads
\be \label{hapa}
ds^2= (1+\frac{r^2}{l^2}) dT^2 - \frac{dr^2}{ (1+\frac{r^2}{l^2})}- r^2
d\Omega^2_{p-2}
\ee
These coordinates are actually the best adapted to write down the would-be
newtonian potential (this is only a somewhat formal concept here in the
 sense that
although the space can be considered static with Killing vector
$\frac{\pd}{\pd T}$, it is not asymptotically flat).
\be
V(r) =\frac{r^2}{l^2}
\ee
which clearly puts into evidence the {\em confining character}
 of the AdS space.
\par

In spite of the fact that we have already introduced coordinates which 
fully cover the space, in some physical applications the AdS space appears
linked to some specific set of coordinates. Let us quickly list some
of the most important ones.

A different, but closely related set of coordinates is the one used by
Susskind and Witten
in \cite{susskind}(see also \cite{gibbons}). The metric has the form

\begin{equation}\label{sw}
ds^2 = \frac{l^2}{(1 - r^2)^2}( -4 \sum_{i=1}^{p-1}
(dx^i)^2 +(1 + r^2)^2 d\tau^2)\, .
\end{equation}

They are easily obtained from the globally static form by

\begin{equation}
\sinh{\chi} = \frac{2r}{1- r^2}\, .
\end{equation}

$CAdS$ itself corresponds to the ball $r < 1$, and the boundary sits on
the sphere $r = 1$.

Another interesting set of coordinates (common to all constant curvature spaces)
is Riemann's, in which the metric reads

\begin{equation}\label{tres}
ds^2 = \frac {\eta_{\m\n}dy^{\m}dy^{\n}}{(1 - \frac{r^2}{4 l^2})^2}\, ,
\end{equation}

\noindent where $\m ,\n= 0,\ldots p-1$ and $\eta_{\m\n}$ is the ordinary Minkowski metric,
and $r^2 \equiv \eta_{\m\n}y^{\m}y^{\n}$.

In order to understand them, it is useful to introduce first another canonical
set of coordinates, valid for any constant curvature space as well (cf. \cite{synge}).
Let us start from the fact that the geodesic
deviation between neighboring geodesics grows as 
\be
\eta = l(\frac{d\eta}{ds})_{s=0} |\sinh \frac{s}{l}|,
\ee
(easily obtained from the geodesic deviation equation).
Now, we use the fact that 
 the angle
between the tangents to such neighboring geodesics is precisely the
volume element on the unit Minkowskian sphere, which can be easily
obtained in terms of the ordinary volume on the unit Euclidean sphere:
$d\Omega^2_{p-1}(hyperbolic)\equiv - d \xi^2 - \sinh^2 \xi
d\Omega^2_{p-2}$.
\par
In that way we can use Pithagoras'theorem of the triangle with hypotenuse 
$ds$ and other sides $dr$ and $d\eta$,
getting easily for the volume element:

\begin{equation}\label{hyperbolic}
ds^2 = dr^2 - l^2 \sinh^2 \frac{r}{l} (d \xi^2 + \sinh^2 \xi d\Omega^2_{p-2})\, .
\end{equation}
\par

\par

Then, Riemann's coordinates can be constructed
(\cite{synge}) by 
\begin{equation}\label{uno}
y^{\m} \equiv 2 l u^{\m} \tanh \frac{r}{2l}\, ,
\end{equation}

\noindent where $u^{\m}$ is the unit tangent vector to the geodesic going 
to the point P from a fiducial point $P_0$; and $r$ is the geodesic
distance from $P_0$ to $P$. Plugging equation (\ref{uno}) into (\ref{hyperbolic})  easily yields (\ref{tres}).
\par

The Poincar\'e coordinates (also called {\em horospheric coordinates} 
in the old british literature)  
\cite{gubser1}  are defined as

\begin{equation}
\left\{
\begin{array}{rcl}
X^0 &=& t/z\, ,\\
& & \\
X^a &=& x^a /z\, ,\\
& & \\
X^p - X^{p-1}& =& \frac{l}{z}\, ,\\
\end{array}
\right.
\end{equation}

\noindent (where $a = 1,\ldots,p-2 $), and the coordinate $z$ is dimensionless.

The metric reads

\begin{equation}\label{horospheric}
ds^2 = \frac{1}{z^2} (dt^2 - d\vec{x}_{p-2}^2 - l^2 dz^2)\, ,
\end{equation}

\noindent where $d\vec{x}_{p-2}^2$ is the Euclidean line element in 
$\mathbb{R}_{p-2}$.

The metric above enjoys a manifest $O(1,p-2)$ symmetry. Besides, it is
invariant under {\em dilatations} $x^{\m}\rightarrow \lambda x^{\m}$
($ x^{\m} = (t,\vec{x},z)$) and {\em inversions} $x^{\m}\rightarrow
\frac{x^{\m}}{x^2}$. Of course those transformations just convey an
action of $O(2,p-1)$ on the horospheres.\footnote{The coordinates used by 
Maldacena in \cite{malda} are essentially $v \equiv
\frac{l}{z}$:

\begin{equation}
ds^2 = - \frac{l^2}{v^2} dv^2 + \frac{v^2}{l^2} dx_{\parallel}^2\, ,
\end{equation}

\noindent (where $dx_{\parallel}^2$ stands for the ordinary Minkowski
 metric in $M_{p-1}$)
\par 
Actually what Maldacena did is to work with the adimensional radius,$\bar{l}$,
(which just happens to be equal to $\bar{l}\equiv (4\pi g N)^{1/4}$),
\be
l^2 \equiv \bar{l}^2 \a'
\ee
where $\a' \equiv\frac{1}{l_s^2}$ is the string tension,
and define a coordinate with mass dimension one by
\be
u\equiv \frac{v}{\a'}
\ee
This yields:
\begin{equation}
ds^2 = \a'[ - \frac{\bar{l}^2}{u^2} du^2 + 
\frac{u^2}{\bar{l}^2} dx_{\parallel}^2 ]\, ,
\end{equation}
}

 Poincar\'e coordinates break down at $z = \infty$ (u = 0), (which we
 shall call the {\em horizon}); which in terms of the embedding is
 just $X^p = X^{p-1}$. In terms of the global static coordinates of
 (\ref{avis}), this equation has solution for a given $\t$ for all $\chi <
 \chi(\t)\equiv \sinh^{-1} |\tan \t |$.
\par
In terms of the coordinates in (\ref{avis}), this means
\be
n^{p-1}=\frac{sin\t}{sin\rho}
\ee
(which has a physical solution as long as $sin\t\leq sin\rho$).
\par

This region can be easily parametrized, using $(X^0)^2 - \d_{ij}X^i
X^j = 1$ $(i = 1\ldots p-2)$ by $X^0 = \cosh z; X^i = n^i \sinh z$,
and the induced metric on the horizon is:

\begin{equation}
ds^2 = -dz^2 - \sinh^2 z d\Omega^2_{p-3}\, .
\end{equation}

The boundary of $AdS$ can be identified with the surface $z=0$, which in
terms of the embedding coordinates is equivalent to $X^p - X^{p-1} =
\infty $.  The normal vector to the boundary is the spacelike vector
$ n=\frac{l^2}{z^2}\frac{\pd}{\pd z}$, which is precisely the reason why we say
that the boundary is a timelike surface. In the coordinates of equation 
(\ref{avis}) it
corresponds to $\rho=\pi/2$, unless $\sin{\t} = \sin{\rho}\, n^{p-1}$ (that is,
the point is in the horizon), in which case a $0/0$ ambiguity is encountered,
and further expansion is needed. 
\par
A glance at equation (\ref{horospheric}) clearly shows that the induced metric
on the boundary is conformal (with a singular factor) to the Minkowski metric.
\newpage

\section{Some Comments on De Sitter space}
The De Sitter space $dS_p$ is in many senses an analytic continuation of
$AdS_p$ . There are, however, two important differences: on the one hand,
 it does not have a timelike infinity; on the other, any observer in it 
has got a horizon. 
\par
It can be globally defined (\cite{Spradlin} ) as the
hypersurface
\be
X_0^2 -\delta_{ij} X^i X^j= - l^2
\ee
($i,j,\ldots = 1\ldots p$),
in a convenient Minkowski space $\mathbb{R}_{1,p}$
\be
ds^2=d X_0^2 -\delta_{ij} d X^i d X^j
\ee
Global coordinates can be defined through
\bea\label{global}
X^0&&= l \sinh{\tau}\nonumber\\
X^i&&= l n^i\cosh{\tau}
\eea
and the metric reads
\be
ds^2 = l^2 (d\tau^2 - \cosh^2{\tau} d\Omega_{p-1}^2)
\ee
The Ricci tensor corresponds to a constant curvature space,
\be
R_{\m\n}=-\frac{p-1}{l^2}g_{\m\n}
\ee
If we perform the change
\be
\cosh{\tau}=\sec{T}
\ee
(where $-\pi/2\leq T \leq \pi/2$)
then the metric reads
\be
ds^2= l^2 \sec^2{T}(dT^2- d\Omega_{p-1}^2)
\ee
This clearly shows that the only natural definition of infinity in $dS_p$ is
at $\tau=\infty$, that is $T=\pm\pi/2$,(a spacelike surface)
 with induced metric corresponding to the  unit 
sphere $S^{p-1}$. This, in turn, does imply that there are horizons associated
to a given observer. The {\em future event horizon} is the boundary between 
events 
which can eventually be detected by the observer and those that can not. 
The {\em past
horizon} is the boundary between those events which can detect the observer,
 and 
those that can not.
\par
The final set of coordinates we will introduce is the so called {\em static}
ones, in which
\bea
&&X^0= l\sqrt{1-r^2} \sinh{ t}\nonumber\\
&& X^a= l\, r\,  n^a\nonumber\\
&&X^p = l \sqrt{1-r^2}\cosh {t}
\eea
(where $a=1\ldots p-1$).
The metric reads
\be
ds^2= l^2( (1-r^2)dt^2-\frac{dr^2}{1-r^2}-r^2 d\Omega_{p-2}^2)
\ee
(which is the analytic continuation to imaginary radius of the $AdS_p$ metric
in (\ref{hapa})). 
\par
By redefining coordinates
\bea
&&\tilde{r}\equiv l r\nonumber\\
&&\tilde{t}\equiv l t
\eea
we get directly the corresponding newtonian potential,
\be
V(\tilde{r}) = - \frac{\tilde{r}^2}{l^2}
\ee
which is a repulsive one.
\par
The Killing vector defining staticity, namely
\be
k=\frac{\pd}{\pd t}
\ee
is not globally timelike; actually it is so only in a wedge covering a 
quarter of $dS_p$, namely $r< 1$.
\par
If we perform a Wick rotation in the global coordinates, (\ref{global}) 
and we want 
that the euclidean manifold is an sphere $S^p$, the timelike angular 
coordinate 
must be periodic,
\be
\tau\sim\tau + 2\pi
\ee
which means (cf. \cite{Birrell}) that the period in the proper length is  
\be
l_0\sim l_0 + 2\pi l
\ee
signalling the presence of a temperature associated to the horizon,
\be
\beta = 2\pi l
\ee

\newpage
\section{ Penrose's Conformal Infinity}

In $AdS_p$ the infinity with the explicit coordinates introduced in
(\ref{avis}) in this sense is located at $\rho = \pi/2$,
so that the metric at the boundary is given by
\be
ds^2 = d\tau^2 - d\Omega_{p-2}^2
\ee
conveying a topological $\mathbb{R}\times S^{p-2}$ structure.

Penrose's construction of conformal infinity (cf. \cite{penrose}) was originally designed 
as a general procedure to address the physics of the asymptotic structure of the space
time in General Relativity. The main idea used to bring infinity back to a finite distance
is to Weyl rescale the physical metric, looking for a new manifold 
$\tilde{\mathcal{M}}$ with boundary, such that the interior of $\tilde{\mathcal{M}}$
coincides with our original spacetime $\mathcal{M}$, endowed with the metric $\hat{g}_{\m\n}$
to be defined in a moment.
In the physics literature it is costumary to denote  the part of $\partial\mathcal{M}$ 
corresponding to end-points of null geodesics by $\mathcal{J}$
\par
Based on the previous construction we will say that a spacetime $(\mathcal{M},g_{\m\n})$ is asymptotically  Einstein 
if there exists a smooth manifold $\tilde{\mathcal{M}}$, with metric $\hat{g}_{\m\n}$, and
a smooth scalar field $\Omega(x)$ defined in $\mathcal{M}$ such that:
\par
i) $\mathcal{M}$ is the interior of $\tilde{\mathcal{M}}$ .
\par
ii)$\hat{g}_{\m\n} = \Omega^2(x) g_{\m\n}$
\par
iii) $\Omega(x) = 0$  if  $x\in\mathcal{J}$; but $N_{\m}\equiv - \nabla_{\m}\Omega$ is nonsingular
in $\mathcal{J}$.
\par 
iv)Every null geodesic in $\mathcal{M}$ has two endpoints in $\mathcal{J}$.
\par
and, finally, the field equation:
\par
v)$R_{\m\n}\sim \lambda g_{\m\n}$
(cf. \cite{penrose}\cite{ashtekar}\cite{hawking}).
\par
(where we  have followed Penrose's notation $a\sim b$ to indicate two things that are
equal on $\mathcal{J}$ only).
\par
For example,  $AdS_p$ in the coordinates used in equation (\ref{avis})
is conformal to  
the metric of Einstein's static universe (ESU). We see that in this example
\be
\Omega\equiv cos \rho
\ee
and the {\em infinite} $\scri$ is located in these coordinates at the finite distance $\rho = \pi/2$.
\par
One of the simplest ways of characterizing the behavior of the conformal factor $\Omega$ is
to study null geodesics in its vicinity (cf. \cite{penrose}). We shall choose an affine parameter $\hat{u}$ on them,
(that is, $\hat{l}^{\m}\hat{\nabla}_{\m}\hat{ u} \sim 1$; where $\hat{l}^{\m}$ is the tangent (null) vector
to the geodesic)  and  we further fix the origin of the affine parameter by $\hat{u}\sim 0$.
\par
There is a corresponding parameter $u$  associated to the metric $g_{\m\n}$, defined by
\be
\frac{d\hat{u}}{d u} \equiv l^{\m}\nabla_{\m}\hat{u}\equiv \Omega^2 \hat{l}^{\m}\hat{\nabla}_{\m}\hat{u} = \Omega^2
\ee
\par
The conformal factor will have, by analiticity, an expansion of the type
\be\label{om}
\Omega(\hat{u}) = -\sum_{n=1}^{\infty}A_n \hat{u}^n
\ee
If $\mathcal{M}$ is an Einstein space with scalar curvature given by
\be
R = \frac{2 n \lambda}{n - 2}
\ee
it is not difficult to show that the vector $N_{\m} \equiv - \nabla_{\m} \Omega$ obeys
\be
\hat{N}^2 \sim \frac{2\lambda}{(n - 1)( n - 2)}
\ee
conveying the fact that the sign of the cosmological constant is related to the spacetime properties of
the $\scri$ boundary (and, in particular, in the ordinary $(1,3)$ case, we see that
 $\mathcal{J}$ will be timelike when the cosmological constant is negative only).
\par
It is also possible to show that
\be
\hat{\nabla}_{\n}\hat{\nabla}_{\m}\Omega = \frac{1}{n} \hat{\Delta} \Omega ~\hat{g}_{\m\n}
\ee
which in turn, (using the fact that $l^2 = 0$), enforce $A_2 = 0$  in the preceding expansion (\ref{om})(\cite{penrose}).
\par
Let us remark that we are here associating to a metric
in $\mathcal{M}$, a whole conformal class in $\mathcal{J}$.(That is, all the construction above is
invariant under $\Omega \rightarrow t~ \Omega$, with $t\in \mathbb{R}^{+}$).
\par
%
\par

\newpage
\section{Witten's Holography}
\subsection{The bulk action expressed in terms of boundary values.}

The special properties of $AdS_p$ allow for uniqueness of a solution
of a wave equation, given data on the boundary. Precise mathematical theorems
can be found in \cite{graham}. Let us concentrate in the
euclidean case, where $\pd EAdS_p = S_{p-1}$.
\par
Using horospheric coordinates (\ref{horospheric}) 
it can be easily shown that the appropiate
Green function is:
\be
K(z,\vec{x},\vec{x}')\equiv l^{p-1}\frac{\Gamma(p-1)}{\pi^{(p-1)/2}\Gamma((p-1)/2)}
\frac{z^{p-1}}{[l^2 z^2 + (\vec{x}-\vec{x}')^2]^{p-1}}
\ee
which obeys the equation
\be
\Delta K(z,\vec{x},\vec{x}')= 0
\ee
as well as
\be
lim _{z\rightarrow 0} K(z,\vec{x},\vec{x}')=  \delta^{p-1} (\vec{x}-\vec{x}')
\ee
Using that, it is plain that the solution for the bulk field $\Phi(z,\vec{x})$,
with action
\be
S\equiv \frac{1}{2}\int_{AdS_p} d(vol) \pd_{\m} \Phi \pd^{\m}\Phi
\ee
and fixed value at the boundary,
\be
\Phi(z=0,\vec{x})=\phi(\vec{x})
\ee
is given by:
\be
\Phi(z,\vec{x})=\int_{S_{p-1}} d(vol)_{\vec{x}'}K(z,\vec{x},\vec{x}')
\phi(\vec{x}')
\ee
In order to compute the bulk action corresponding to fixed boundary values,
it is convenient to first regularize the boundary to $z=\epsilon$
and only at the end make $\epsilon = 0$. In that way one gets:
\be
S(\phi)\sim - \frac{p-1}{2}\int d(vol)_{\vec{x}}d(vol)_{\vec{x}'}\ 
l^{p-2}\ \frac{\phi(\vec{x})\phi(\vec{x}')}{(\vec{x}-\vec{x}')^{2(p-1)}}
\ee
which is equivalent to the fact, on the CFT side
\be
<\mathcal{O}(\vec{x})\mathcal{O}(\vec{x}')>\sim
\frac{1}{(\vec{x}-\vec{x}')^{2(p-1)}}
\ee
\par
In the massive case, we have to add to the action an extra term:
\be
\Delta S\equiv \frac{1}{2}\int d(vol) m^2 \Phi^2
\ee
The wave equation is now best analyzed in the euclidean version
of the coordinates (\ref{hyperbolic}), namely
\be
ds^2 = dy^2 +l^2 \sinh^2 (y/l) ~d\Omega_{p-1}^2
\ee
For large values of the coordinate $y$, the laplacian on the sphere
should be negligible, and the equation reduces to:
\be
e^{-(p-1)y/l}\frac{d}{dy}(e^{(p-1)y/l}\frac{d}{dy}\Phi)=m^2\Phi
\ee
which admits an exponential behaviour $e^{\lambda_{\pm}y/l}$, with
$\lambda_{\pm}(\lambda_{\pm} + p-1)=l^2 m^2$. This means that in the massive 
case it is not possible to extend to the bulk an arbitrary function $\phi$
on the boundary. Massive field should tend to boundary fields coupling to
operators $\mathcal{O}_{\Delta}$ with scale dimension $(p-1) + \lambda_{+}$,
because  a boundary Weyl transformation $y\rightarrow y + w$
can be compensated by the transformation $\phi\rightarrow 
e^{- \frac{w}{l}\lambda_{+}}\phi$, which means that the boundary fields have got,
 in this sense, scale dimension $-\lambda_{+}$.
\par
The Green's function is now:
\be
K(z,\vec{x},\vec{x}')\equiv l^{p-1+2\lambda_+}\ \frac{\Gamma(p-1+\lambda_{+})}{\pi^{(p-1)/2}
\Gamma((p-1)/2+\lambda_{+})}
\frac{z^{(p-1)+\lambda_{+}}}
{(l^2 z^2+(\vec{x}-\vec{x}')^2)^{(p-1)+\lambda_{+}}}
\ee
leading easily to the bulk action:
\be
S\sim\int d(vol)_{\vec{x}}d(vol)_{\vec{x}'}\ l^{p-2+2\lambda_+}\ \frac{\phi(\vec{x})\phi(\vec{x}')}
{(\vec{x}-\vec{x}')^{2(p-1+\lambda_{+})}}
\ee
Given the fact that $\lambda_{\pm}\equiv -\frac{p-1}{2}\pm \frac{1}{2}
\sqrt{(p-1)^2+4 l^2 m^2}$, the scaling dimensions of the operators which
can be represented in this way are necessarily $\Delta > \frac{p-1}{2}$
\par 
It can be further shown (cf. \cite{klebanovwitten},\cite{muck})
that the use of the {\em irregular boundary conditions} 
($\lambda_{-}$) allows
to obtain correlators for CFT operators with scaling dimensions 
$(p-1)/2-1<\Delta<(p-1)/2$.

\subsection{Operator Mapping}
Let us examine the massless fields of the $IIB$ string theory living in
 $AdS_5\times S_5$. Besides the field corresponding to the graviton, 
$g_{\m\n}$, which will be expanded around the background as $\bar{g}_{\m\n}
+ h_{\m\n}$, there is a complex scalar $B$, a complex two-form, 
$A^{(2)}_{\m\n}$ and a real self-dual four form, $A^{(+)}_{\m\n\rho\sigma}$.
The fermionic sector consists in a complex gravitino $\psi_{\m}$, as
well as a complex fermion
$\lambda$.
\par
All these fields are expanded (cf. \cite{Kim}) in terms of spherical 
harmonics corresponding
 to the sphere $S_5$, so that for example in the scalar case
\be
\bar{g}^{ab}h_{ab}=\sum \pi^{I_1}Y^{I_1}
\ee
Antisymmetric tensors need more terms in the expansion:
\be
A_{\m\n}=\sum a^{I_{10}}Y^{I_{10}}_{[\m\n]}+\ldots
\ee
\par
On the other hand, the different fields of ${\cal{N}}=4$ SYM can be packed 
in several ways (cf. \cite{Kovacs}). For example, in terms of ${\cal{N}}=1$ 
superfields, they span three chiral superfields $\phi^I$ (transforming 
on the adjoint of the gauge group) as well as a vector superfield $V$, 
transforming also in the adjoint. The three scalar superfields give three 
complex scalars and three Weyl fermions. The vector superfield give another
 Weyl fermion, namely the gaugino, and a real vector. In this language, 
only a $SU(3)\times U(1)$ subgroup of the full $SU(4)$ symmetry is manifest.
In terms of ${\cal N}=2$ superfields, everything can be packed into a 
vector plus a hypermultiplet. In the vector there is a complex scalar (which we shall call $a$), the 
vector field an a couple of fermions; and in the hypermultiplet there are 
four real scalars
and another two fermions.
\par
Witten (\cite{witten}) gave the first entries of a dictionary relating 
fields on the two sides of the correspondence. Let us consider, for
 example,  the 
${\cal{N}}=1$ superfield 
$T^{I_1\ldots I_n}\equiv tr (\phi^{I_1}\ldots\phi^{I_n})$, which
has got dimension $n$, which corresponds to a conformal weight $\lambda=n-4$.
Looking at the mass formula, this means that the corresponding bulk field 
has a mass
$m^2=n(n+4)$. This corresponds in the $IIB$ side to the expansion of the 
graviton trace.
\par
Another field is 
$V^{I_1\ldots I_n}\equiv tr (W_a W^a \phi^{I_1}\ldots\phi^{I_n})$
where $W_a$ is the superfield strengh, and the term $W_aW^a$ 
contains a gluino bilinear. Its total dimension is $n+3$, so that the mass of the bulk field is $m^2= (n+3)(n-1)$. This corresponds to the expansion of the two-form in the $IIB$ theory.
\par
Finally, there is the field $Q^n\sim tr( a^{n-2}F_{\m\n}F^{\m\n}+\ldots)$
(where $a$ is the particular scalar included in the ${\cal{N}}=2$ 
vector multiplet). 
Its dimension is $n+2$, so that the mass of the corresponding bulk field is 
$m^2=(n+2)(n+6)$. This corresponds to the expansion of the traceless graviton 
in the $IIB$ theory.
\par
All these operators enjoy special properties that guarantee protection of 
their dimensions from quantum corrections.

\subsection{Finite Temperature}
Following the holographic philosophy, if we want to represent a conformal theory
 at finite temperature (which in the euclidean case means that we are working
 in a manifold $M_n=S_{n-1}\times S_1$ or $M_n=\mathbb{R}_{n-1}\times S_1$,
 the first thing we have to do is to look for a negative curvature manifold $B_{n+1}$ such that
$\pd B_{n+1}=M_n$. It so happens that Hawking and Page in \cite{hawkingpage}
 studied this very problem, and discovered that there are two such manifolds.
The first one is essentially EAdS with time running in a circle:
\be
ds^2= (1+\frac{r^2}{l^2}) dT^2 + \frac{dr^2}{ (1+\frac{r^2}{l^2})}+ r^2
d\Omega^2_{p-2}
\ee
with $T=T+\beta^{\prime}$, where $\beta^{\prime}$ is in principle arbitrary.
\par
The second manifold is Schwarzschild Anti-de Sitter, which we will call
 SAdS. Its metric is:
\be
ds^2= (1+\frac{r^2}{l^2}-\frac{c_n M}{r^{n-2}}) dT^2 +
\frac{dr^2}{ (1+\frac{r^2}{l^2}-\frac{c_n M}{r^{n-2}})}+ r^2
d\Omega^2_{p-2}
\ee
where the constant $c_n=\frac{16\pi G\Gamma(n/2)}{(n-1)2\pi^{n/2}}$.
The {\em horizon} is defined by 
\be
(1+\frac{r^2}{l^2}-\frac{c_n M}{r^{n-2}})|_{r=r_{+}}=0
\ee
When we compactify the euclidean time on a circle there will in general appear
 a conic singularity, unless the temperature happens to have the particular value:
\be
\beta_0\equiv\frac{4\pi r_{+} l^2}{n r_{+}^2+(n-2)l^2}
\ee
The topology of SAdS is $\mathbb{R}^2\times S_{n-1}$. It is possible to define
 a rescaling such that the topology is $\mathbb{R}^2\times \mathbb{R}^{n-1}$
namely,
\bea
&&\rho^n\equiv \frac{b^{n-2}}{c_n M} r^n\nonumber\\
&&t^n\equiv\frac{b^{n-2}}{c_n M}\tau^n
\eea
Then, in the high mass limit, $M\rightarrow\infty$ we can neglect the $1$ in
 the metric coefficients, getting
\be\label{hr}
ds^2= (\frac{\rho^2}{b^2}-\frac{b^{n-2}}{\rho^{n-2}})d\tau^2 +\frac{d\rho^2}{
(\frac{\rho^2}{b^2}-\frac{b^{n-2}}{\rho^{n-2}})}+ \rho^2 d\vec{x}^2
\ee
The radius of the sphere $S^{n-1}$ is now of the order $M^{1/n}$, so that in
 the lint the topology is as stated, $\mathbb{R}^2\times
 \mathbb{R}^{n-1}$.This solution had been previously considered by Horowitz
 and Ross in \cite{horowitzross}.
\par
It is natural to assume that when there are several manifolds $B_{n+1}$ 
 bounding the
same $M_n$, one should consider a superposition of the two. In a given
physical situation, the dominant contribution will be provided by the
solution with least action.
In our case the action is given by
\be
I\equiv-\frac{1}{2\kappa^2}\int \sqrt{g}d^{n+1}x(l-2\lambda )=\frac{n}
{\kappa^2 l^2}V_{n+1}
\ee
York's boundary term  vanishes, and the second equality stems from the
 fact that the scalar curvature is $R=2\frac{n+1}{n-1}\lambda$. We have represented by $V_{n+1}$ the
 total volume of the space, which diverges. If we regurarize through a cutoff
$\epsilon^{-1}$, then
\bea
&&Vol(AdS)=\int_0^{\beta^{\prime}}dt\int_0^{\epsilon^{-1}}dr r^{n-1}\int _{S^{n-1}}d\Omega \nonumber\\
&&Vol(SAdS)=\int_0^{\beta_0} dt\int_{r_{+}}^{\epsilon^{-1}}dr r^{n-1}\int
 _{S^{n-1}}d\Omega 
\eea
We can determine $\beta^{\prime}$ by demanding that the geometry of the
 hypersurface $r=\epsilon^{-1}$ is the same both in AdS and in SAdS. This means
 that:
\be
\beta^{\prime}\sqrt{1+\frac{r^2}{l^2}}=\beta_0\sqrt{1+\frac{r^2}{l^2}-\frac{c_n
 M}{r^{n-2}}}|_{r=\epsilon^{-1}}
\ee
This gives
\be
\beta^{\prime}=\beta_0 (1-\frac{1}{2}c_n M l^2 \epsilon^n)
\ee
yielding for the difference in action the value:
\be
\Delta I\equiv \frac{n}{\kappa^2} lim_{\epsilon\rightarrow 0}(I(SAdS)-I(AdS))=
\frac{Vol(S_{n-1})}{\kappa^2}\frac{\pi r_{+}^{n-1} l^2}{n r_{+}^2+(n-2)l^2}
(l^2 - r_{+}^2)
\ee
The average value of the energy is given by:
\be
< E > =\frac{\pd I}{\pd \beta_0}= \frac{(n-1)r_{+}^{n-2}
 Vol(S_{n-1})}{2\kappa^2}(r_{+}^2 + l^2)
\ee
In such a way that the canonical entropy reads:
\be
S\equiv \beta_0 < E > - \Delta I = \frac{Vol(2\pi l^2 r_{+}^{n-1}
 S_{n-1})}{\kappa^2}\sim A
\ee
where $A$ is the area of the horizon.
\par
All this is consistent with the AdS/CFT conjecture. When $\beta_0\rightarrow 0$
one expects the hight temperature limit on the CFT side, which means that the
entropy density should behave as
\be
S\sim T^{n-1}
\ee
On the gravity side,
\be
r_{+}=\frac{n-2\beta_0}{4\pi^2}
\ee
which we discard, or else
\be
r_{+}=\frac{4\pi l^2}{n \beta_0}
\ee
This last possibility gives
\be
S_{AdS}\sim \beta_0^{-(n-1)}
\ee
\newpage
\section{Wilson Loops}
It is natural , following Maldacena, to make the ansatz that the value of the
Wilson loop $C$ (supposedly lying on the four fimensional submanifold $z=0$)
is given in the leading approximation by the area of the minimal area surface
$D$ (such that $C=\pd D$) extending on
the five-dimensional bulk manifold. Before proceeding, it is easy to show that
in AdS, conformal invariance prevents an area law, because if we start from
the expression
\be
W(C)=e^{- A(D)}
\ee
and  rescale by $\lambda$, then 
\be
W(\lambda C)=e^{-A(\lambda D)}=e^{-A(D)}
\ee
(by dilatation invariance of AdS). This implies that
\be
W(\lambda C)=W(C)
\ee
This clearly leads to
\be
A\sim \frac{T}{L}
\ee
and to an expression for the static potential
\be
V\sim\frac{1}{L}
\ee
It is worth remarking that this argument ceases to apply in the SAdS
case. Using the blowup
\be
ds^2= (\frac{\rho^2}{b^2}-\frac{b^{n-2}}{\rho^{n-2}})d\tau^2 +\frac{d\rho^2}{
(\frac{\rho^2}{b^2}-\frac{b^{n-2}}{\rho^{n-2}})}+\rho^2 d\vec{x}^2
\ee
(where now the loop itself is placed at $\rho=\infty$) we see that
\be
\rho\geq b
\ee
so that the presence of the horizon breaks conformal invariance and allows
the possibility of confining behavior.
\par
Let us see in some detail how the calculation proceeds in the conformal 
situation.
\par
To begin with, there are no quarks in the fundamental in $\cal{N}$=4 SYM.
What we can do is to start with a stack of $N+1$ D-branes, and pull one of
them apart from the others. The long strings stretched between the pack and
the isolated brane reproduce the $W$ boson with a very large mass, which then
behaves in some respects as a particle in the fundamental.
\par
Furthermore, the natural operator to consider (in the sense that this is the
one that comes from dimensional reduction of $\cal{N}$=1SYM in ten
dimensions)
is:
\be
W\equiv\frac{1}{N}tr\, P e^{\int ds( i A_{\m}\dot{x}^{\m}+ \Phi_a \dot{z}^a)}
\ee
(the reason for the funny $i$ is explained in \cite{Drukker})
where $x^{\m}= x^{\m}(s)$ is a parametrized loop in ordinary Minkowski space,
and $ z^a= z^a(s)$ another loop in the complementary six dimensional space.
It can be argued that $\dot{x}^2 =\dot{z}^2$. In the CFT side, this is the
condition for the absence of a linear divergence.
On the string side, there is another reason dealing with boundary conditions.
The total ten dimensional metric can be written in horospheric coordinates as:
\be
ds^2=\frac{1}{z^2}(dx_{1,3}^2-l^2 \delta_{ab}dz^a dz^b)
\ee
where $z^a\equiv z n^a$ and  $\vec{n}$ is a unitary vector living on $S^5$, 
$\vec{n}^2= 1$.
In \cite{Drukker} the following boundary conditions have been proposed on the
imbeddings of the string in the space-time:
\be
x^{\m}(\sigma_1,0)=x^{\m}(\sigma_1)
\ee
and
\be
J_1\,^{\a}\pd_{\a}z^a (\sigma_1,0)=\dot{z}^a(\sigma_1)
\ee
(where $J$ is the two dimensional complex structure on the worldsheet of the
string) and we have parametrized the boundary of the worldsheet as
$\sigma_2=0$. The additional condition that the minimal surface terminates at
the boundary of AdS, $z^a=0$, is only compatible with the above
boundary conditions precisely when  $\dot{x}^2 =\dot{z}^2$. In this case the
boundary conditions can be written as Dirichlet boundary conditions on $S^5$:
\be
n^a(\sigma_1,0)=\frac{\dot{z}^a}{|\dot{z}|}
\ee
\par
All this means that we are really computing:
\be
W(C)\equiv\frac{1}{N} tr\, P e^{i \int ds A_{\m}(\xi)\dot{\xi}^{\m}+\theta^I(s)
X_I(\xi)\sqrt{\dot{\xi}^2}}
\ee
The one-dimensional loop is imbedded into $AdS_5\times S_5$ through
\be
s\in S^1\rightarrow ((\xi^{\m}(s),u(s)),\theta^I(s))
\ee
Let us assume that we place the loop at the boundary, $u(s)=\infty$ and that
besides we map the loop to a fixed point on the sphere,
$\theta^I(s)=\theta^I_0$. The simplest way to proceed (\cite{maldacena}) is to
consider a rectangular static loop, extending from $x=-L/2$ to $x=L/2$,
and in the temporal direction from $0$ to $T$. In that way, in the large T
limit we can easily extract the static potential,
\be
W\sim e^{-T V(L)}
\ee
We can furthermore parametrize the two-dimensional surface bounded by the loop
by $\sigma=x^1\equiv x$ and $\tau= x^0\equiv t$. Assuming that the surface
itself extends only in the holographic direction, and owing to invariance
under time translations, it is uniquely characterized by only one
function
\be
u(x)
\ee
The induced metric on the two-surface is
\be
ds^2=h_{ab}d\sigma^a d\sigma^b=\frac{u^2}{\bar{l}^2}dt^2
+(\frac{u^2}{\bar{l}^2}+(\frac{\bar{l}^2}{u^2}\frac{\pd u}{\pd x})^2) dx^2
\ee
so that its area is given by:
\be
A\equiv \int dt dx \sqrt{det h_{ab}}
\ee
that is
\be
A=T\int_{-L/2}^{L/2}dx \sqrt{\frac{u^4}{\bar{l}^4} +(u_x)^2}
\ee
The problem of finding the minimal area surface bounded by the loop $C$ is equivalent to
minimizing the above expression in terms of the function $u(x)$. This can be
easily done by using the expression for the first integral coming from the
fact that $x$ itself is an ignorable coordinate.
The result is
\be
x=\frac{\bar{l}^2}{u_0}\int_1^{u/u_0}\frac{dz}{z^2\sqrt{z^4-1}}
\ee
Where $u_0$ is the unknown value of the minimum of the funcion $u(x)$). Its
numerical value can be determined by enforcing the boundary condition:
\be
\frac{L}{2}=\frac{\bar{l}^2}{u_0}\int_1^{\infty}\frac{dz}{z^2\sqrt{z^4-1}}
\ee
This gives
\be
A=T u_0\int dy \frac{y^2}{\sqrt{y^4-1}}
\ee
Which, although goes as $T/L$ (because $u_0\sim 1/L$) as it should by conformal invariance, actually
diverges. This divergence can be eliminated by substracting the {\em free
  loop} corresponding to the cuboid extending all
the way to $u=0$:
\be
A_{ren}=T u_0\int dy (\frac{y^2}{\sqrt{y^4-1}}-1)
\ee
\par
The fact that the effective string tension $T_{eff}\sim \lambda^{1/2}$ implies
then that the static potential behaves in this case as:
\be
 V(L)\sim\frac{\lambda^{1/2}}{L}
\ee
\newpage

\section{The Fefferman-Graham construction}


In  mathematical terms the problem associated with the holographic projection
is that of finding the conformal invariants of a given manifold $M_n$, with generic
signature $(p,q)\equiv((1)^p,(-1)^q)$ and dimension $n = p + q$, in terms of the Riemannian invariants 
of some other manifold $\tilde{M}$
in which $M_n$ is  contained in some precise sense.
In order to employ a more physical terminology, we will refer to $M_n$ as the {\em space-time}
and to $\tilde{M}$ as the {\em bulk} (in case it has one dimension more) or {\em ambient space} (in the case it has got two more dimensions).
\par
Following \cite{fefferman} we will work out this geometrical problem from two different points of view, depending
on the dimension and signature of the bulk space. In the so-called {\em Lorentzian } approach the 
ambient space $A_{n+2}$ will have signature
$(p+1,q+1)$ while in the second approach, based on Penrose's definition of
conformal infinity the bulk space $B_{n+1}$ will have
either $(p,q+1)$ or $(p+1,q)$ signature, leading to two different kinematical types of geometric holography.

\subsection{Lorentz Holography}

\par
Conformal invariant tensors , considered as functionals of the metric tensor, $g$, $P(g)$ are defined by the transformation law;
\be
P(\lambda g) = \lambda^{-\Delta} P(g)
\ee
where $\Delta$ is the conformal weight; they are thus associated with a
 given conformal class of metrics,
$[g]$. 
\par
We shall represent the extra two coordinates of the ambient space by $\rho$ and $t$.
Remarkably enough,
Fefferman and Graham were able to prove that diff invariant expressions on
$A_{n+2}$ give rise to conformal invariants
provided that the metric on $A_{n+2}$,  $d\tilde{s}^2$ is such that:
\begin{eqnarray}
i)& d\tilde{s}^2(\rho = 0) = t^2 ds_n^2 \nonumber\\
ii)& d\tilde{s}^2(x,\lambda t,\rho) = \lambda^2 d\tilde{s}^2(x,t,\rho)\nonumber\\
iii)& R_{\m\n}(\tilde{g}) = 0
\end{eqnarray}
where $ds_n^2$ is a convenient reference metric  chosen in $M_n$.
\par
In the odd case, $n\in 2\mathbb{Z}+1$ there is a
perturbative solution to this mathematical problem as a formal series in the
variable $\rho$. Moreover, it so happens that the conditions just stated force the metric in $B$ to be of
the form
\be
ds^2=t^2 ds^2(x,\rho) - 2\rho dt^2 - 2 td\rho dt
\ee
where $ds^2(x,\rho)$ is such that
\be
ds^2(x,\rho=0) = ds_n^2(x)
\ee
which is the fiducial line element in $M_n$
The generator of the dilatations $t\rightarrow \lambda t$ is
\be
T\equiv t\frac{\pd}{\pd t}
\ee
On the region of $A_{n+2}$ defined by $\rho=0$ the following
relationships are true:
\bea
&&\tilde{R}_{abct}=0\nonumber\\
&&\tilde{R}_{abcd}=t^2W_{abcd}\nonumber\\
&&\tilde{R}_{abc\rho}=t^2 C_{abc}\nonumber\\
&&\tilde{R}_{\rho ab\rho}=\frac{t^2}{n-4}B_{ab}
\eea
where the latin indices $a,b,c\ldots\in (1\ldots n)$ and we have explicitly indicated
the extra indices  $t$ and $\rho$. The symbols $W$, $C$ and $B$ stand for
the Weyl, Cotton and Bach tensors, defined
 (in any dimension) by means of the tensor
\be
A_{\a\b}\equiv\frac{1}{n-2}(R_{\a\b} - \frac{R}{2(n-1)} g_{\a\b})
\ee
  as:
\be
W_{\a\b\m\n}\equiv R_{\a\b\m\n}- (A_{\b\m} g_{\a\n} + A_{\a\n}g_{\b\m}  -
A_{\b\n}g_{\a\m}  - A_{\a\m}g_{\b\n})
\ee
(this definition implies that the Weyl tensor  vanishes identically when n=2
or n=3). The Weyl tensor is conformal invariant of weight $\Delta=-1$.
When $n>3$ the space is conformally flat iff $W=0$.
\par

On the other hand, the Cotton tensor, $C_{\m\n\rho}$ is  defined by:
\be
C_{\a\b\g}\equiv \nabla_{\a}A_{\b\g} - \nabla_{\b} A_{\a\g}
\ee
For dimension $n=3$ the Cotton tensor is a conformal invariant of weight
zero. For bigger dimension, it is not conformal invariant.
In n=3 the spacetime is conformally flat iff the Cotton tensor vanishes

and
\be
B_{\m\n}\equiv \nabla^{\rho}C_{\rho\m\n} + A^{\a\b}W_{\a\b\m\n}
\ee

It is to be stressed that the Bach tensor is conformally
invariant (of weight $\Delta=1$)  when n=4 {\em only}.
\par
Let us consider the simplest  example in order to visualize this result. 
We shall take $M=S^1$. The defining equation
\be
x_1^2+x_2^2=1
\ee
can equally well be written in projective coordinates
$x^i\equiv\frac{\xi^i}{\xi^0}$($i=1,2$) as the cone $\mathcal{C}$:
\be
\xi_0^2 -\xi_1^2 - \xi_2^2 = 0.
\ee
In this picture the  ambient space $A_3$ is then defined in terms
of those (projective) coordinates $\xi^0,\xi^1,\xi^2$.
\par
 The dilatation generator on the  cone 
$\mathcal{C}$, that is, the vector
\be
T\equiv \xi^{\m}\frac{\pd}{\pd \xi^{\m}}
\ee
is forced to be a null vector with respect to the ambient metric $\tilde{g}$ 
(restricted to $\mathcal{C}$). All this is obviously equivalent to
identify the cone $\mathcal{C}$ with the {\em light cone} of the three-dimensional ambient space with signature $(1,2)$
(where the coordinate $\xi_0$ is a {\em time}); that is,
\be
d\tilde{s}^2= d\xi_0^2-d\xi_1^2-d\xi_2^2
\ee
In terms of the variables $(\xi_0,x_1,x_2)$ this reads
\be 
d\tilde{s}^2= (1-x_1^2-x_2^2)d\xi_0^2-\xi_0^2(dx_1^2 +
dx_2^2)+2\xi_0d\xi_0(x_1 dx_1 + x_2 dx_2)
\ee
so that on $\mathcal{C}$
\be
d\tilde{s}^2|_{\mathcal{C}}= -\xi_0^2(dx_1^2 +dx_2^2)
\ee
Canonical ambient coordinates can be introduced by first passing to polar
coordinates $(\xi_0,r,\theta)$, with
\bea
&&\xi_1=r\cos{\theta}\nonumber\\
&&\xi_2=r\sin{\theta}
\eea
in such a way that 
\be
d\tilde{s}^2=d\xi_0^2-dr^2-r^2d\theta^2
\ee
and then defining the {\em holographic coordinates}
\bea
&&t\equiv 2 (\xi_0 + r)\nonumber\\
&&\rho\equiv\frac{1}{2}\frac{r-\xi_0}{r+\xi_0}
\eea
yielding the metric in the canonical form:
\be
d\tilde{s}^2= - t^2 \frac{(\rho + 2)^2}{4}d\theta^2- 2\rho dt^2-2t dt  d\rho
\ee
We clearly see that we have changed the signature from $(0,1)$ in $M=S^1$ to
$(1,2)$ in the ambient space.
\par
Generically, one goes from signature $(p,q)$ in $M$, to signature
$(p+1,q+1)$ for the Lorentzian ambient space; that is, of the two extra
coordinates, one is spacelike and the other timelike.
\par
In the even case, $n\in 2\mathbb{Z}$, there is an obstruction to the
perturbative
solution, the Fefferman-Graham tensor, $F_{\m\n}$, which is nothing other than
the Bach tensor when $n=4$, but in general dimension is a new tensor, a
conformal invariant of weight $(n-2)/2$. As we shall see later on, this
obstruction is related to the conformal anomaly.

\subsection{Penrose holography}
There is another, mathematically equivalent construction, based in a bulk space,
$B_{n+1}$ such that $M_n=\pd B_{n+1}$ in Penrose's sense. If we extend the
coordinates of $M$, $x^a$ by a new holographic coordinate, $r$, such that
$r=0$ is precisely the boundary, then the construction is such that the metric
in the bulk space obeys:
\be\label{penrose}
ds_{-}^2=\frac{1}{r^2}(- dr^2+ g^{-}_{ab}(x,r)dx^a dx^b)
\ee
(we shall see the reason for the $-$ label in a moment).
The bulk space is a constant curvature space, that is
\be
R^{-}_{ab}=n g^{-}_{ab}
\ee
(in our previous conventions, this is equivalent to normalize the total radius
of AdS to unity, $l = 1$). The signature of this metric, as advertised, is $(p,q+1)$.
\par
There is a canonical way of constructing the bulk metric from the ambient metric.
If we consider the hypersurface in the ambient space defined by constant
values of the modulus of the dilatation generator,
\be
T^2= -1
\ee
that is, $2\rho t^2=1$.  This space of codimension one, which we shall
identify with the bulk space, $B_{n+1}$, contains $M_n$ as its conformal
boundary. The bulk metric as induced by the imbedding, is given by
\be
ds^{(-)2}_{n+1}=\frac{1}{2\rho}ds^2(x,\rho) - \frac{1}{4\rho^2}d\rho^2
\ee
(that is, the extra coordinate is spacelike). This can be put
precisely of the
form in (\ref{penrose}) 
\be
ds^{(-)2}_{n+1}=\frac{1}{r^2}(-dr^2+ds^2(x,\rho=\frac{r^2}{2}))
\ee
through $r^2\equiv 2\rho$.

\par
\par
It should be clear by the  reasoning above that when $M$ is of signature
 $(p,q)$, $B_{n+1}$ enjoys signature $(p,q+1)$.

\par
We could  repeat the previous construction for 
\be
T^2=1
\ee
with the the result
\be
(ds^{(+)2})_{n+1}=-\frac{1}{2\rho}ds_{-}^2(x,\rho) - \frac{1}{4\rho^2}d\rho^2
\ee
(This is actually the promised reason for the subscript $\pm$ on the metric).
For example, in our simplest $M=S^1$ case, this gives
\be
ds^{(+)2}_{n+1}=\frac{t^2 (\rho+2)^2}{8\rho}d\theta^2 - \frac{1}{4\rho^2}d\rho^2
\ee
the extra coodinate is timelike now. The metric can be formally put into the
canonical form through $\rho=-\frac{r^2}{2}$:
\be
ds^{(+)2}_{n+1}=\frac{1}{r^2}(-dr^2 + ds^2(x,\rho=\frac{-r^2}{2}))
\ee
\par
In general this would mean that the holographic coordinate is in this case timelike, and therefore
in order to get a boundary with the desired signature $(p,q)$ we should consider a bulk space of signature $(p+1,q)$ (instead
of $(p,q+1)$).
\par
The preceding remark is potentially interesting for a {\em boundary} physical spacetime of Minkowskian
 signature $(1,3)$, since in this case we could try to perform a holographic projection with positive cosmological constant,
but on a bulk spacetime with signature $(2,3)$. This could presumably be the
 mathematical basis for the de Sitter/CFT duality proposed by Strominger in \cite{strominger}.
\par
\subsubsection{Appendix}
There is a useful generalization of the usual horospheric coordinates which
 gives the metric induced on pseudospheres by the imbedding on a flat space of
 arbitrary signature. 
 Actually, for arbitrary $\pm$ signs, denoted by $\epsilon_i = \pm 1$, the metric induced on the surface
\be
\sum_{i = 1}^n  \epsilon_i x_i^2 = 1
\ee
by the imbedding on the flat space with metric
\be
ds^2 =\sum_{i = 1}^n  \epsilon_i d x_i^2 
\ee
can easily be reduced to a generalization of Poincar\'e's metric for the half-plane by introducing the coordinates
\bea
&&z\equiv x^{-}\nonumber\\
&&y^{\m}\equiv z ~x^{\m}
\eea
where we have chosen the two last coordinates, $x^{n-1}$  and $ x^n$ in such a way that their contribution to the metric
is $ dx_{n-1}^2 - dx_n^2$ (this is always possible if we have at least one timelike coordinate); and we define
$x^{-}\equiv x^{n} - x^{n-1}$. $\m\in (1,\ldots n-2)$.
The generalization of the Poincar\'e metric is:
\be
ds^2 = \frac{\sum \epsilon_{\m} dy_{\m}^2 - dz^2}{z^2}
\ee
\subsection{The News Function and the Holographic Map}
The geometric holography just described strongly depends on the existence of a unique solution for a Cauchy problem
defined in terms of Einstein's equations on the bulk and with initial conditions fixed by the conformal class of the physical
spacetime metric at the boundary. A necessary condition for the uniqueness of 
the solution is the vanishing of the
(gravitational) Bondi-Sachs news function $N$ through $\mathcal{J}$ (cf. (\cite{alvarez})). 
The precise definition of the complex quantity $N$ is:
\be
N\equiv -\frac{1}{2}R_{\m\n}\bar{m}^{\m}\bar{m}^{\n}
\ee
where $m^{\n}$ is one of the elements of a complex null Newman-Penrose tetrad,
which is formed by four null vectors, two of them real, $l^2=0$ and $n^2=0$,
and two complex conjugate of one another, $m^2=0$ and $\bar{m}^2=0$. They are
normalized in such a way that $ l.n = - m.\bar{m} = 1 $.
\par

The physical meaning of this is that the spacetime boundary
$\scri$ is {\em opaque} to gravitational radiation; in the four-dimensional case, with topology $S^3\times \mathbb{R}$
absence of Bondi-Sachs news requires that the Bach tensor vanishes on $\scri$. 
To be specific, (cf. \cite{penrose}) the {\em variation} of  Bondi mass is given by an integral of
two terms. The first one is a convenient projection of the energy-momentum
tensor of the matter, whereas the second one is proportional to the modulus of
the news function:
\be
\delta M=\int_{\Sigma} A^2 T_{\m\n}n^{\m}n^{\n} + \frac{N\bar{N}}{4\pi G}
\ee
where $A$ is a scalar fied defined asymptotically in terms of the formerly
introduced vector $N_{\m}\equiv -\nabla_{\m}\Omega$ by
\be
\hat{N}^{\m}= A n^{\m}
\ee
and $\Sigma$ is a surface comprised between two cuts of the conformal
boundary, $\scri$.
\par
When n=3 (with topology $S^2\times \RR$), the Bach tensor vanishes, conveying the fact that there are no gravitational
news in this case (\cite{ashtekar}). This is the simplest instance of the much
alluded to general theorem proved in \cite{fefferman} staying that for a 
spacetime $M$ of even dimension  there is no obstruction for the existence of a formal power series solution to the
Cauchy problem with initial data on the boundary.
\par
In the case of a n=3 boundary spacetime $S^2\times \RR$, however, Bondi news exist in general for matter fields if the Cotton tensor
does not vanish. This means that for four dimensional bulk spaces there is the possibility of having a well defined Cauchy problem
in the Fefferman-Graham sense, and yet, Bondi news for fields with spin different from 2. It is obvious that this is problematic from the
holographic point of view (except in the case of pure gravity).
\par
Geometrically,  the vanishing of the Cotton tensor in the three-dimensional case is the necessary and
 sufficient condition for the existence of conformal Killing spinors. (In the four dimensional case, the equivalent condition 
(implying that the space is conformally Einstein)  is
 the vanishing of the Bach tensor (cf.\cite{kozameh}). Only in this case the conformal symmetry is realized asymptotically
in such a way that one can define asymptotically conserved charges associated to the $O(3,2)$ conformal group ($O(4,2)$ in
the four-dimensional case).
\par 
This reduction of the asymptotic symmetry group to AdS is similar to the reduction from the asymptotic Bondi-Metzner-Sachs
\cite{penrose} group in the asymptotically flat case, towards Poincar\'e, as has been pointed out in \cite{ashtekar}. In this case, however,
the condition $B_{\m\n}\sim 0$ is too strong, and, in particular, it is not stable against gravitational perturbations.
There is then a curious discontinuity in the limit $\lambda\rightarrow 0$.
\par
It then would seem that the vanishing of the conformal anomaly in the three dimensional case
in the holographic setting \cite{henningson} does not need the vanishing of the Cotton tensor.
\par
Another fact worth stressing is that gravitational Bondi news are generically non-vanishing when $\scri$ is spacelike,
or even null. It is plain that the interplay between holography and Bondi news is related to the existence of a Cauchy surface 
for asymptotically anti de Sitter spacetimes (i.e. $\scri$ timelike).
The simplest example is, obviously, AdS itself, where in order to define a Cauchy surface one is forced to impose 
{\em reflective} boundary conditions on $\scri$,\cite{avis}
 enforcing the desired absence of Bondi news for matter fields..
\par
Remarkably enough, Hawking \cite{hawking} has proved that the physics of this  set of boundary conditions
is equivalent to assuming that the gravitational fields tend to AdS at infinity fast enough.
Physically, absence of Bondi news on $\scri$ is necessary in order that a CFT living on $\scri$
could {\em propagate} holographically to the bulk in a unique way.

\newpage

\section{Holography and the Conformal Anomaly}

As we have just seen, in the framework of the geometric approach to holography  in its Poincar\'e form (that is, when the holographic image $M_d$ is
represented as Penrose's conformal infinity of another $B_{d+1}$ manifold),
there is a privileged system of coordinates such that the {\em boundary}
$\pd B_{d+1}\sim M_d$ is located at $\rho = 0$, namely
\be\label{canon}
ds^2 = \frac{l^2 d\rho^2}{4\rho^2} + \frac{1}{\rho} h_{ij}(x,\rho)dx^i dx^j
\ee\label{hs}
(This coordinate is related to the canonical one in (\ref{penrose}) by $\rho=r^2$ ; the normalization corresponds to a cosmological constant 
$\lambda\equiv -\frac{d(d-1)}{2 l^2}$ when $h_{ij}
=\d_{ij}$).
Physically, the boundary condition is
\be
h_{ij}(x,\rho = 0) = g_{ij}(x)
\ee
where $g_{ij}$ is an appropiate metric on $M_d$.
\par
Those coordinates are in conformal backgrounds essentially our old friends the horospheric coordinates: $z\equiv \rho^2$.)

\par
The Ricci tensor for the metric (\ref{canon}) can be expressed as
\bea
&&R_{\rho\rho}=-\frac{d}{4\rho^2}+\frac{1}{4}tr( h^{-1}h^{\prime})^2-
\frac{1}{2}tr (h^{-1}h^{\prime\prime})\nonumber\\
&&R_{\rho i}=\frac{1}{2}\nabla_j (h^{-1}h^{\prime})^{j}_i-\frac{1}{2}\nabla_i(tr
h^{-1}h^{\prime})\nonumber\\
&&R_{ij}=R_{ij}[h]-\frac{2-d}{l^2}h^{\prime}_{ij}-\frac{2
  \rho}{l^2}h^{\prime\prime}_{ij}-\frac{d}{\rho l^2}h_{ij}+
\frac{1}{l^2}tr(h^{-1}h^{\prime})h_{ij}\nonumber\\
&&-\frac{\rho}{l^2}tr(h^{-1}h^{\prime})h^{\prime}_{ij}+\frac{2\rho}{l^2}(h^{\prime}h^{-1}h^{\prime})_{ij} 
\eea
where a prime means $\frac{d}{d\rho}$,and $\nabla_i$ is the covariant 
derivative
of the Levi-Civita connection of the metric $h_{ij}$.

Einstein's equations
\be
R_{\m\n}=- \frac{d}{l^2} g_{\m\n}
\ee
can be then rewritten for the metric(\ref{canon}) as:
\bea\label{einstein}
\rho [2 h''_{ij} - 2 h'_{il}h^{lm}h'_{mj} + h^{kl}h'_{lk} h'_{ij}]- 
l^2 R_{ij}-
(d-2)h'_{ij} - h^{kl}h'_{kl} h_{ij} &=&0\nonumber\\
(h^{-1})^{jk}(\nabla_i h'_{jk} - \nabla_k h'_{ij}) &=&0\nonumber\\
(h^{jk}h''_{kj}) -\frac{1}{2} (h^{il}h'_{lm}h^{mn}h'_{ni})&=&0
\eea
\par

There is a natural scale symmetry associated with the preceding metric,
namely
\bea
&&\rho\rightarrow \lambda \rho\nonumber \\
&& h_{ij}\rightarrow\lambda h_{ij}
\eea
The famous Fefferman-Graham obstruction implies, however, that , when $d\in
2\mathbb{Z}$, there appear logarithmic terms in the expansion
of the preceding metric around $\rho = 0$, which begin at $\rho^{d/2}$, and spoil
a consistent power solution. (Although they are absent if one uses dimensional
regularization, as in \cite{imbimbo},\cite{mazur}). As has been shown by Henningson and Skenderis in
\cite{henningson} (following a suggestion of E. Witten in \cite{witten}), these terms are
responsible for the conformal anomaly. Let us sketch their argument. In the
 basic work by Fefferman and Graham it is proved that there is a formal power
series solution to Einstein's equations with negative cosmological constant,
up to $\rho^{d/2}$. This means that in $d=4$, for example,
 a consistent expansion exists
of the form
\be\label{metric}
h_{ij} = g_{ij} +h^{(1)}_{ij} \rho  + h^{(2)}_{ij}\rho^2  + 
\tilde{h}^{(2)}_{ij} \rho^2 \log{\rho} + o(\rho^3)
\ee
Even the term $h^{(d/2)}_{ij}$ is not completely determined; Einstein's
equations only give its trace as well as its covariant
derivative.(cf. \cite{haro}).  
\par
It is, on the other hand, obvious that the Einstein-Hilbert action is
divergent. To be specific,

\be
S\equiv \frac{1}{2\kappa_{d+1}^2}\left[\int_{M_{d+1}} d(vol)_{d+1} (R_{d+1} - 
2 \lambda_{d+1}) + \int_{M_d} d x_d 2 K\right]
\ee
where $K\equiv h_{(ind)}^{ij}\nabla_i n_j$ is the trace of the second fundamental form, $n^i$
being the normal to the boundary and $h_{(ind)}^{ij}\equiv h^{ij}-n^i n^j$ the
induced metric on the boundary.
\par
Explicit calculation shows that $R-2 \lambda = -\frac{4}{1-d}\lambda 
= -\frac{2d}{l^2}$
\bea
L_{bulk}&=&-\frac{2d}{l^2}\int 
\sqrt{\frac{l^2}{4 \rho^2}\rho^{-d}}h^{1/2}d\rho\nonumber\\
&=&-\frac{d}{l}\int \rho^{-1-d/2}h^{1/2}d\rho.
\eea
This integral is, as advertised, divergent, a reflect of the fact that AdS is a non 
compact space. Actually, it diverges in both limits, both $\rho=\infty$
(which should correspond to the infrared (IR) region in the CFT, according to
the IR/UV connection) and in $\rho=0$, which is the UV region of the CFT.
The divergence at $\rho=\infty$ would appear only  at order $\rho^{d/2+1}$ or higher
in the expansion of the metric, which is higher that the order that can be
determined unambiguosly.
\par
Let us concentrate in the UV divergences.
A way to regularize them is to cut-off the integral over
$\rho$ with a $\theta(\rho - \epsilon )$. This leads to an inverse power 
series in $\epsilon$
\be\label{reg}
S_{(\epsilon)}\sim \sum_{n = d/2}^{0} \epsilon^{- n}S^{(n)} +\hat{S} \log{\epsilon} + 
S_{ren} 
\ee
The logarithmically divergent term comes from the integral of the 
$\rho^{d/2}$ term in the
expansion of the $M_d$ volume element, combined with the pre-factor
$\rho^{- d/2 - 1}$. This explains why it only appears for even $d$. It is 
remarkable that this term has {\em a priori} nothing to do with the logarithmic ambiguity
of the expansion noticed above, (although cf. later on) and is a purely bulk effect.
\par
Actually, if we write the regularized action in the form,
\be
S_{(\epsilon)}\equiv \frac{1}{2\kappa_{d+1}^2}\int \sqrt{g}dx_d L_{\epsilon}
\ee
then
\be
L_{\epsilon}=a_0 \epsilon^{-d/2} + a_1\epsilon^{-d/2+1} +\ldots + 
\epsilon^{-1} a_{d-1}-\log~\epsilon~a_d + L_{finite}
\ee
Incidentally, the two logarithmically divergent terms in equations
(\ref{metric}) and (\ref{reg}) are related in the sense
that, as has been proved in \cite{haro},
\be
\tilde{h}^{d/2}_{ij}=-\frac{4}{d\sqrt{g}}\frac{\d}{\d g^{ij}}\int dx_d \sqrt{g}\, a_{(d)}
\ee
Now, under the scale invariance mentioned above, all powers are invariant
 by themselves (cf(\cite{henningson})) , meaning that the variation of the logarithmically divergent
 term has to be cancelled with an anomalous variation of the finite part:
if $\lambda=1+2\delta \sigma$, then $\delta h_{ij}=2 h_{ij}\delta\sigma$ and
$\delta(\log{\epsilon})=2\delta\sigma$
\be
 - \delta S_{ren}\equiv  \int_{M_d} \sqrt{g} dx_d \delta \sigma \cal{A} 
\ee
Where the {\em anomaly}, $\mathcal{A}$ is given by:
\be
\mathcal{A}= - \frac{a_{(d)}}{\kappa_{d+1}^2}
\ee
General theorems \cite{bonora} ensure that the anomaly can
 always be written as:
\be
a_d = d~l^{d-1} (E_d + I_d + D_i[g] J^i_{d-1})
\ee
where $E_d$ is proportional to Euler's density in $d$ dimensions, $I_d$
is a conformal invariant, and the total derivative can be cancelled by 
a (finite covariant) counterterm.
\par
In \cite{henningson} this property has been used to compute the Weyl anomaly 
in several interesting cases, by just expanding carefully $h^{1/2}$, and
 finding complete agreement for the leading term when $N\rightarrow \infty$.
\par
For example, in the physically important case of $d=4$, the logarithmically divergent terms read:
\be
S=\frac{2}{l \kappa_{d+1}^2}\int d^4 x \sqrt{g} \ log~\epsilon \ [1/2 (g^{ij}h_{(2)ij})
-1/4  (g^{il}h_{(1)lm}g^{mn}h_{(1)ni}) + 1/8 (g^{ij}h_{(1)ji})(g^{kl}h_{(1)kl})]
\ee
The $\rho^0$ term of the first of Einstein's equations (\ref{einstein})
gives:
\be
R_{ij}=-\frac{1}{l^2} ( 2 h_{(1)ij} + g^{kl}h_{(1)kl} g_{ij})
\ee
so that
\be
R = -\frac{6}{l^2}  g^{ij} h_{(1)ji}
\ee
and (using $h^{ij}= g^{ij}-\rho g^{il}h_{(1)lm}g^{mj} + \rho^2 
(g^{il}h_{(1)lm}g^{mn}h_{(1)np}g^{pj} - g^{il}h_{(2)lm}g^{mj})$),
\be
R^{ij}R_{ij}=\frac{1}{l^4} (4  g^{il}h_{(1)lm} g^{mn} h_{(1)ni} + 8  (g^{il}
h_{(2)li})^2)
\ee
whereas the third yields:
\be
 g^{ij}h_{(2)ji} = \frac{1}{4}  g^{il}h_{(1)lm} g^{mn} h_{(1)ni}
\ee

This altogether leads to:
\be
a_4 = \frac{l^3}{8}(-R^{ij}R_{ij} + \frac{1}{3}R^2)
\ee
The four dimensional invariants are:
\be
E_4\equiv \frac{1}{64} (R^{ijkl}R_{ijkl} - 4 R^{ij}R_{ij} + R^2)
\ee
and
\be
I_4= - \frac{1}{64}(R^{ijkl}R_{ijkl} - 2 R^{ij}R_{ij} + \frac{1}{3} R^2)\equiv
W^{ijkl}W_{ijkl}
\ee
(where $W_{ijkl}$ is the Weyl tensor).
\par
The anomaly is then given by
\be
\mathcal{A}=\frac{1}{2\kappa^2_5}(-2a_4)= - \frac{N^2}{\pi^2}(E_4+I_4)
\ee
where we have used the fact that
\be \frac{1}{\kappa^2_5} = \frac{Vol(S_5)}{\kappa^2_{10}}
=\frac{l^5 \pi^3}{64\pi^7 g_s^2 l_s ^8}
\ee

and that

\be
l = (4\pi g_s N)^{1/4} l_s 
\ee
This reproduces the leading term in the large $N$ limit of the four dimensional
conformal anomaly, which is given in full by:
\be
\mathcal{A}= - N^2\frac{1 -N^{-2}}{\pi^2}(E_4+I_4)
\ee
\par
In (\cite{blau}) non-leading (in $N$) contributions to the 
Weyl anomaly were computed,
with only partial success.
\par

Incidentally, for any Ricci-flat metric on $M_d$, Einstein's equations 
for $M_{d+1}$
are obeyed with
\be
h_{ij}(x,\rho) = g_{ij}(x)
\ee

\subsection{PBH Diffeomorphisms}
The Penrose-Brown-Henneaux (PBH),(\cite{penrose}\cite{brown}\cite{imbimbo}) 
diffeomorphisms were introduced in \cite{imbimbo} as particular bulk 
diffs which
include conformal transformations on the boundary.
\par
If we impose in the canonical form of the bulk metric we employed in equation 
(\ref{canon}) that
the diff is such that
\be
\delta g_{d+1,d+1}=\d g_{d+1,i}=0
\ee
(with $x^{d+1}\equiv\rho$), then we get that the diff must be generated by a
vector such that
\bea
\xi^{n+1}&&=-2\rho\sigma(x)\nonumber\\
\xi^i&&=a^i (x,\rho)
\eea
and, besides,
\be\label{isty}
\pd_{\rho}a^i=-\frac{l^2}{2} h^{ij}\pd_j \sigma
\ee
This implies, in particular, that
\be\label{metrica}
\d h_{ij}= \,^{(h)}\nabla_i \xi_j+\,^{(h)}\nabla_j \xi_i-2\sigma h_{ij}-2\sigma\rho\pd_{\rho}h_{ij}
\ee
We assume that there is an analytic expension
\be
a^i=\sum_{n=1}a^i_{(n)}\rho^n
\ee
which implies that, to the lowest order in the holographic coordinate,the diff
is a pure Weyl transformation on the boundary metric defined on
$M$,$h_{ij}^{(0)}\equiv g_{ij}$
\be
\d g_{ij}=-2\sigma g_{ij}
\ee
(where as in the last paragraph, we assume an expansion $h_{ij} =\sum_{q}h^{(q)}_{ij}\rho^q$). 
\par
The variation of the other terms in the expansion are easily obtained from
(\ref{metrica}). For example, the next one is:
\be
\d h^{(1)}_{ij}=\,^{(0)}\nabla_i a^{(1)}_j+\,^{(0)}\nabla_j a^{(1)}_i
\ee
\par
The basic differential equation just written down in eq. (\ref{isty})
determines the different terms in the expansion of the PBH diffs in terms of
the coefficients in the expansion of the bulk metric. For example, the first
one is:
\be
a^i_{(1)}=\frac{l^2}{2} g^{ij}\pd_j \sigma
\ee

Imbimbo et al first noticed in \cite{imbimbo} the remarkable fact 
that from these variations
it is easy to get expressions for the coefficients in the expansion of the
bulk metric, for example,
\be
h^{(1)}_{ij}=\frac{l^2}{d-2}(R_{ij}-\frac{1}{2(d-1)}R g_{ij})
\ee
This formula fails if the spacetime dimension is $d=2$; this illustrates the
claims made in the last paragraph on $h_{ij}^{(1)}$. In this case, for
example,
\be
h^{(1)}_{ij}=\frac{1}{2}(R g_{ij}+ t_{ij})
\ee
with $\nabla_i t^{ij}=0$ and $g^{ij}t_{ij}=-R$.

\par

Sometimes there are terms which appear with arbitrary coefficients; this
phenomenon starts at second order in which
\be
c_1 l^4 W_{klmn}W^{klmn}g_{ij}+ c_2 W_{iklm} W^{klm}_j
\ee
can be added to the expression of $h^{(2)}_{ij}$ for any
$c_1$ and $c_2$. 
\par
Were not for these constants, this procedure would allow to determine the {\em
  bulk} metric in terms of boundary data; that is, to {\em decode the 
hologram}. 
\par
What is perhaps even more remarkable is that the whole approach can be used to
recover the conformal anomaly in any dimension.
\par
In order to achieve this goal,we shall consider an arbitrary gravitational
action in the bulk space. We shall only assume that the Dirichlet problem for
the metric has a unique solution.
If we chose to write the action in the form (\cite{imbimbo})
\be
S\equiv\frac{1}{2\kappa_{d+1}^2}\frac{l}{2}\int d\rho dx_d \rho^{-(1+d/2)}
\sqrt{g(x)} b(x,\rho)
\ee
(where we assume that $b$ is a functional of $g$ on shell), then by expanding
on a power series in the holographic coordinate
\be
b(x,\rho)\equiv\sum_n b_n(x)\rho^n
\ee
and performing the integration over $d\rho$, one gets
\be
S=\frac{1}{\kappa_{d+1}^2}\sum_{p\neq d/2}\frac{1}{2 p -d}\int d^d x \sqrt{g} b_d(x).
\ee
There is a pole in the expansion for any even dimension. Actually, the
coefficient $b_p$ represents a trace anomaly in dimension $d=2p$.
\par

Then, using the fact that the total variation of the integrand $I$ under any
 diff generated by the vector $\xi$,
must be
\be
\d I = \nabla_{\a}(\xi^{\a} I)
\ee
as well as the curious property that for PBH diffs (where $\xi=a^i\pd_i-2\rho\sigma\pd_{\rho}$),
\be
\d \sqrt{h}=\nabla_{\mu}\xi^{\m}=\,^{(h)}\nabla_i a^i + d\sigma -\rho\sigma
h^{-1}\pd_{\rho}h
\ee
and also that PBH act as Weyl on the boundary,
\be
\d \sqrt{g}=d\sigma\sqrt{g}
\ee
then, the PBH variation of the construct b is easily found to be:
\be
\d b=-2\sigma\rho\pd_{\rho}b+\,^{(0)}\nabla_i(b a^i)
\ee
which can be easily translated in corresponding formulas for the modes $b_p$,
 for example
\be
\d b_0=0
\ee
\be
\d b_1=-2\sigma b_1 + \frac{l^2}{2} b_0\Box\sigma
\ee
These formulas start having arbitrary parameters in $\d b_3$, reflecting the
 corresponding arbitrariness in $h^{(2)}_{ij}$. The authors of \cite{imbimbo}
 have argued that from here, local expressions for the modes
 can always be found. The first two are:
\be
b_0=constant
\ee
\be
b_1=b_0\frac{l^2}{4(d-1)}R
\ee
Starting with $b_2$ there is an increasing number of arbitrary parameters in
 the solution. It is however possible to use this information to find the form
 of the Euler density contribution to the conformal anomaly valid for any
 gravitational action with the characteristics indicated (\cite{imbimbo}).

\subsection{The Holographic Energy-momentum tensor}
The expectation value of the boundary energy momentum tensor 
(cf. \cite{balasubramanian},\cite{emparan},\cite{kraus},\cite{cappelli}) 
is given by the
variation of the gravitational action with respect to the metric on the
 boundary (cf. \cite{brown} for a comprehensive treatment of related matters).
\par

The starting point is the regularization of the gravitational action we made
earlier in (\ref{reg}):
\be
S_{(\epsilon)}=\frac{1}{2\kappa^2}\int d x_d \sqrt{g}\left[ \epsilon^{-d/2}a_{(0)}+
\epsilon^{-d/2 + 1}a_{(2)}+\ldots +\epsilon^{-1}a_{(d-2)} - \log{\epsilon}
a_{(d)}\right] +S_{ren}
\ee
where $S_{ren}$ defines the {\em renormalized action}.
\par
The expectation value of the energy momentum tensor is:
\be
<T_{ij}>=\frac{2}{\sqrt{g}}\frac{\d}{\d
  g^{ij}}S_{ren}=lim_{\epsilon\rightarrow 0} \epsilon^{1- d/2} T_{ij}[\gamma]
\ee
where $T_{ij}[\gamma]$ is the energy-momentum tensor corresponding to the
regulated theory with respect to the induced metric on the boundary, 
$\gamma_{ij}\equiv\frac{1}{\epsilon}h_{ij}(x,\epsilon)$ (cf. \cite{haro}).
\par
This energy momentum can, in turn, be separated in two different
contributions, coming from the regulated action, and from the counterterms:
\be
T_{ij}[\gamma]=T^{\epsilon}_{ij}[\gamma]+T^{counterterms}_{ij}[\gamma]
\ee
Haro et al have given in ref. \cite{haro} explicit expressions for this energy
momentum tensor in different dimensions. In the simplest of all cases, $d=2$,
one gets
\be
<T_{ij}>=\frac{l}{2\kappa^2}t_{ij}=\frac{l}{\kappa^2}(h^{(2)}_{ij}-g_{ij}g^{lm}h^{(2)}_{lm})
\ee
And indeed we recover in that way the standard two-dimensional conformal anomaly:
\be
<T^i_i> = -\frac{l}{2\kappa^2}R
\ee

Let us remark, finally, that the explicit transformation rules under PBH
diffs, combined with those formulas, allow to determine the explicit  Weyl variations of
the holographic energy momentum tensor.
For example, in the much discused two-dimensional example:
\be
\d <T_{ij}>=\frac{l}{2\kappa^2}(\nabla_i\nabla_j\sigma - g_{ij}\nabla^2\sigma)
\ee

\newpage

\newpage
\section{Conclusions}
In a sense, mathematical holography is pure kinematics. Is this all there is
to it?
\par
In spite of much effort devoted to it, the extension of the above ideas to
non conformal situations is still unclear.
\par
One topic which seems worth exploring is to clarify the r\^ole of the cutoff
(cf. \cite{Gubser}) and its suggested relationship with the 
Randall-Sundrum approach (\cite{randall}).
\par
A curious, but well-known fact is that all black hole solutions of the 
holographic type dominate the path integral
(that is, enjoy lower action) only in the region in which the specific heat is
positive, $c_V>0$. This leaves open the question as to whether holography is
possible at all for systems such as the Schwarzschild black hole, for which
the specific heat is always negative.
\par
It has been recently suggested that some of these ideas could be extended to
the constant positive curvature  spaces (de Sitter, dS)(\cite{strominger}), by
analizing the asymptotic diffeomorphisms in the Brown and Henneaux sense
(cf. \cite{brown}). It remains to explore in detail its physical meaning as well as how these ideas fit in the 
Fefferman-Graham
framework.

\newpage

\section*{Acknowledgments}
We have benefited from many discussions with Luis \'Alvarez-Gaum\'e, 
C\'esar G\'omez , Juan Jos\'e Manjar\'{\i}n and Tom\'as Ort\'{\i}n. 
This work ~~has been partially supported by the
European Commission (HPRN-CT-200-00148) and CICYT (Spain). L.H. has been supported by a MCT grant AP99 .      



\end{document}